# Feasibility study of online tuning of the luminosity in a circular collider with the robust conjugate direction search method *

JI Hong-Fei[1;3], JIAO Yi[2;3;1)], WANG Sheng [1;3], JI Da-Heng[2;3], YU Cheng-Hui[2;3], ZHANG Yuan[2;3], HUANG Xiao-Biao[4]

1 (Dongguan Institute of Neutron Science (DINS), Dongguan 523808, China)

2 (Key Laboratory of Particle Acceleration Physics and Technology，Institute of High Energy Physics, Chinese Academy of Sciences)

3 (Institute of High Energy Physics, Chinese Academy of Sciences, Beijing 100049, P.R. China)

4 (SLAC National Accelerator Laboratory, Menlo Park, California 94025, USA)

**Abstract:** The robust conjugate direction search (RCDS) method has high tolerance to noise in beam experiments. It has been demonstrated that this method can be used to optimize the machine performance of a light source online. In our study, taking BEPCII as an example, the feasibility of online tuning of the luminosity in a circular collider is explored, through numerical simulation and preliminary online experiments. It is shown that the luminosity that is artificially decreased by a deviation of beam orbital offset from optimal trajectory can be recovered with this method.

## 1 Introduction

A particle accelerator is typically a complex system that consists of many components, e.g., beam transport, control, diagnostic and acceleration systems. There are many variables to be tuned to achieve good machine performance. Therefore, optimization usually needs to be performed in a multi-dimensional variable space, at every stage of the construction of an accelerator: design, commissioning, and operation.

In the design phase, a variety of programs (e.g., Mad [1], SAD [2], Elegant [3], AT [4]) and algorithms (e.g., genetic algorithms [5], multi-objective genetic algorithm [6]) have been developed to model the system and to further optimize the expected machine performance. However, during commissioning or operation of an existing accelerator, optimization of the machine performance is not as easy as in the design phase. There inevitably exist various kinds of errors which lead to some differences between the theoretical model and the actual conditions. As a result, the optimal values of the variables derived from theoretical analysis and numerical calculation may not work well during real operation. Actually, in most cases, physicists usually do optimization in a manual manner, repeatedly tuning and scanning variables around the design variables according to the main performance parameters observed directly from the machine. Although this method usually works, it is time-consuming and its effectiveness decreases with increasing number of variables.

It is noted that, along with the unceasing development of computer technology and optimization algorithms, online optimization of accelerator performance has become both imperative and feasible. Several algorithms have been proposed for different purposes in many laboratories; these include slow feedback systems [7], the downhill simplex method [8-10], rotation rate tuning [11], random walk optimization [12], and the robust conjugate direction search (RCDS) method [13]. Of these methods, the RCDS algorithm is effective in optimizing a single-objective function of several variables. It has both high tolerance to noise and high convergence speed. This method has been successfully applied to the SPEAR3 storage ring for real accelerator optimization problems, such as minimization of the vertical emittance with skew quadrupoles and the optimization of the injection kicker bump match [13].

Received date

*Supported by National Natural Science Foundation of China (11475202, 11405187) and Youth Innovation Promotion Association of Chinese Academy of Sciences (No. 2015009)

1) E-mail: jiaoyi@ihep.ac.cn

Based on the success of the RCDS method with a light source, it is interesting to find possible applications of this method in optimizing the performance of a collider. As is known, luminosity is the most important measure of performance of a collider, and can be treated as a single-objective function of several variables. For BEPCII [14], the upgrade project of the Beijing Electron Positron Collider, the luminosity depends on more than 20 variables, including the transverse offset in displacement and angular deviation ($x$, $x'$, $y$, $y'$) at the interaction point (IP), the x-y coupling parameters ($R_1$, $R_2$, $R_3$, $R_4$), working point, RF parameters, and optical parameters at the IP.

To explore the feasibility of using RCDS to online optimize the luminosity of a circular collider, a numerical simulation of luminosity tuning with the RCDS algorithm is first performed and the results are shown in Sec. 2. Then preliminary online experiments of the luminosity tuning in BEPCII are performed, with the results described in Sec. 3. Finally, discussion and concluding remarks are given in Sec. 4.

## 2 Numerical simulation of luminosity tuning with the RCDS method

We apply the RCDS method to luminosity tuning in numerical simulation for two purposes. One is to verify the feasibility of the application in a collider based on simulation results, and the other is to identify some of the possible constraints in experimental applications.

The luminosity is somewhat arbitrarily modeled as a function of $n$ variables ($\eta_1$, $\eta_2$, …, $\eta_n$),

$$Lum(\eta_1,\eta_2,...,\eta_n) = L_{\max}\prod_{i=1}^{n}F(\eta_i), \tag{1}$$

where $L_{\max}$ is the maximum luminosity ($L_{\max}$ is set to 100 cm$^{-2}$s$^{-1}$), and $F(\eta_i)$ describes the dependence of the luminosity on each variable,

$$F(\eta_i) = e^{-\frac{(\eta_i-a_i)^2}{\sigma_i^2}} \frac{J_0(\frac{\eta_i-a_i}{\delta_i})+\Delta_i}{\Delta_i+1}, \tag{2}$$

where a Bessel function is adopted to study the convergence ability and efficiency of the algorithm for a multivariable function with local optima; the global optimal condition is at $\eta_i = a_i$, corresponding to $F(\eta_i) = 1$; $\sigma_i$ determines the decline rate of $F(\eta_i)$ with $|\eta_i\text{-}a_i|$; $\delta_i$ determines the oscillation frequency of the Bessel function; and $\Delta_i$ is set to be larger than 0.5 to avoid minus $F(\eta_i)$. The contributions of coefficients ($\sigma_i$, $\delta_i$, $\Delta_i$) to the function $F(\eta_i)$ are illustrated in Fig. 1.

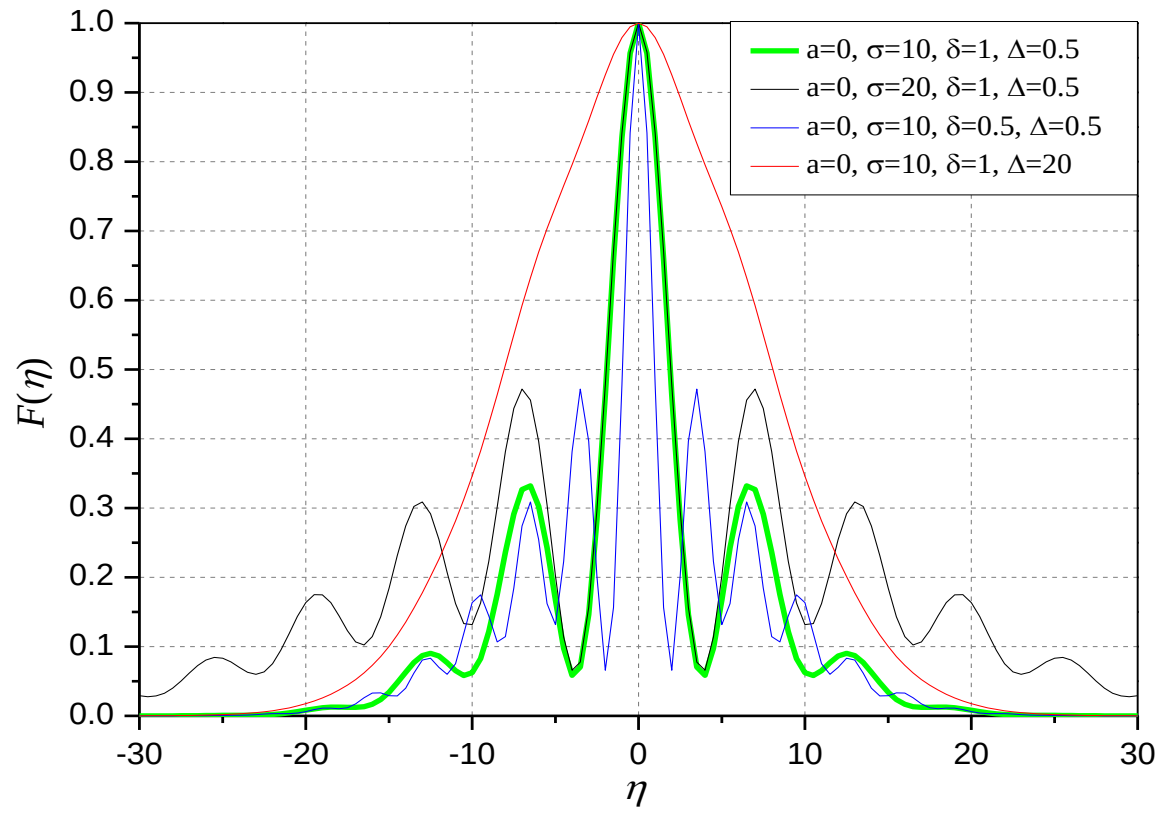

Fig. 1. (color online) Contributions of the coefficients ($\sigma_i$, $\delta_i$, $\Delta_i$) ($i$ = 1, …, $n$) to the function $F(\eta_i)$.

In the numerical experiment, eight variables were considered, i.e., ($x$, $x'$, $y$, $y'$, $R_1$, $R_2$, $R_3$, $R_4$) at the IP. The optimal values of the variables and the coefficients ($\sigma_i$, $\delta_i$, $\Delta_i$) ($i$ = 1, …, 8) were pre-determined based on experience and are listed in Table 1. A number of sets of the variables were randomly generated, and the corresponding luminosity values were evaluated from Eqs. (1) and (2). In each evaluation, a random number (with rms value of 0.5) was added to the calculated luminosity, representing the fluctuation of the luminosity due to various errors in an actual collider. The variables were scanned and optimized with the RCDS algorithm coded in the Matlab program. Fig. 2 shows the objective function for all trial solutions and the best solutions. One can see that for eight variables the luminosity converges to a high value (with a minus sign) after about 300 evaluations. Further study shows that if there are fewer variables, convergence can be reached with fewer evaluations, e.g., 60 evaluations for 4 variables.

Table 1. Coefficients and optimal values of each variable for numerical simulation of luminosity tuning with the RCDS method

| variables ($i$ = 1, …, 8) | $a_i$ | $\sigma_i$ | $\delta_i$ | $\Delta_i$ |
|---|---|---|---|---|
| $\eta_1$ ($x$) | 44.5 μm | 60 μm | 0.3 μm | 10 |
| $\eta_2$ ($y$) | 13.7 μm | 10 μm | 0.2 μm | 0.8 |
| $\eta_3$ ($x'$) | 8.4 mrad | 100 mrad | 0.5 mrad | 15 |
| $\eta_4$ ($y'$) | 2.3 mrad | 100 mrad | 0.5 mrad | 15 |
| $\eta_5$ ($R_1$) | 0.12 | 68 | 0.2 | 20 |
| $\eta_6$ ($R_2$) | 0.07 | 95 | 0.2 | 20 |
| $\eta_7$ ($R_3$) | 0.03 | 68 | 0.2 | 20 |
| $\eta_8$ ($R_4$) | 0.39 | 95 | 0.2 | 20 |

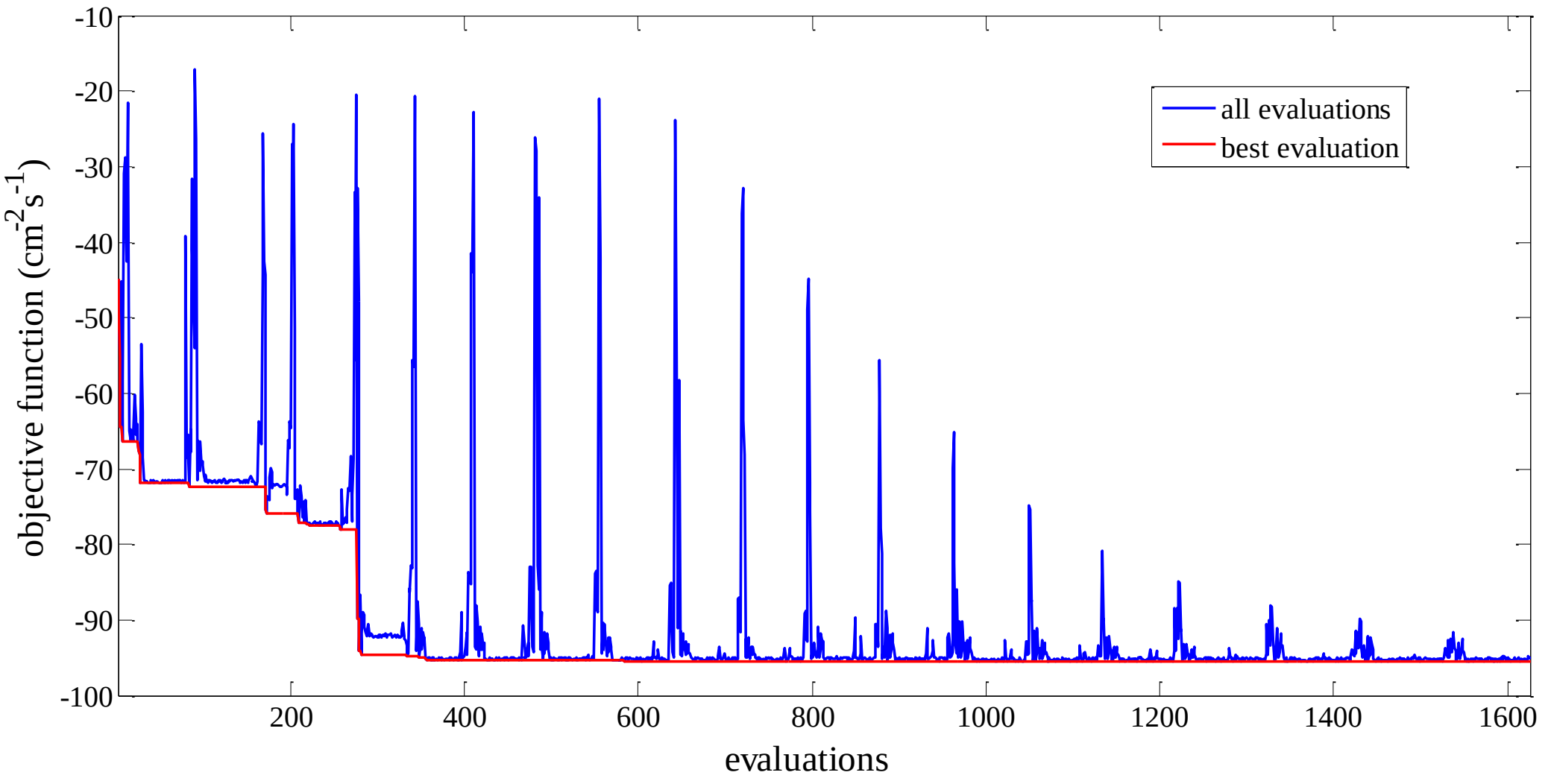

Fig. 2. History of all evaluated solutions and the best-to-date solution during luminosity tuning with RCDS in a numerical simulation.

## 3 Online experiments of luminosity tuning in BEPCII with the RCDS method

To perform online RCDS experiments, the algorithm was coded in SAD, the control program used in BEPCII. The flowchart of the RCDS method and the data stream between the RCDS and operation of BEPCII are shown in Fig. 3. The RCDS method is an optimization algorithm to minimize a multi-variable function. This algorithm iteratively searches all directions in a conjugate direction set. For the search in each direction, a unique line optimizer is applied (see [13] for detail), which is much more robust against noise than traditional algorithms. In the search loop, the direction set is updated according to the criteria on orthogonality and conjugacy. The search loop terminates if the desired convergence or the given number of iterations is reached.

The RCDS is applicable online when the objective function can be measured and the corresponding variables can be tuned online. The program is parameterized by the noise level of the luminosity (the fluctuation due to various errors during operation), the conjugate direction set, a set of variables and their reasonable tuning ranges. In the online application in BEPCII, the set of variables values produced by the RCDS method is set in the collider; then the corresponding luminosity is read back as the objective value and returned to the algorithm for the next iteration. All these actions proceed in a fully automated mode. The iteration continues with data transmission between the algorithm and operation of BEPCII until the luminosity reaches a certain convergence.

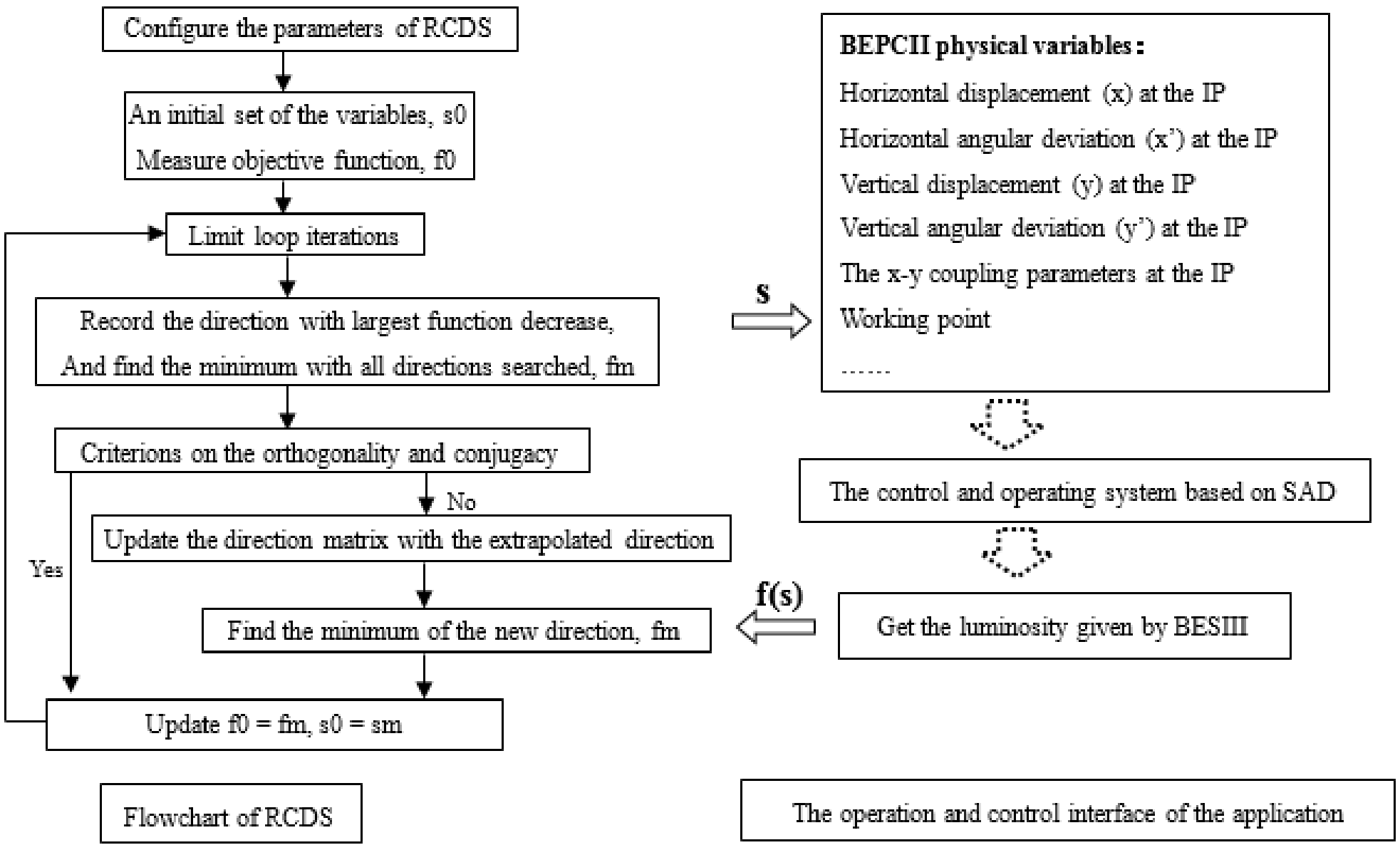

Fig. 3. Flowchart of RCDS for online luminosity tuning in BEPCII, and data stream between the RCDS method and operation of BEPCII.

### 3.1 Safety and efficiency of the machine experiment

As an online application, a rapid and stable running of the RCDS scan is necessary. Thus, several measures have been carried out to enhance the operational speed as much as possible and to improve the ability of the program to handle sudden changes in the state of the collider.

For a fixed number of variables, the runtime of the RCDS program is essentially determined by the time taken for each evaluation. The time for an evaluation can be expressed as

$$t_{evaluation} \approx t_c + t_l + t_v \quad , \tag{3}$$

where $t_c$ represents the time taken for the program to perform the calculation of the values of the variables and the value of the objective function, $t_l$ represents the time taken to get the corresponding luminosity, and $t_v$ represents the time taken for the code to calculate the values of the parameters of the relevant components according to the variable values , to set the parameters automatically by the BEPCII control system and then to wait to obtain a reliable response value of the luminosity. Compared to the others, the time $t_c$ is extremely short, so it can be ignored.

To shorten $t_l$, the objective function was chosen to be the so-called specific small-angle luminosity (SSAL), instead of the BEPCII luminosity as measured by the BESIII End-cap Electromagnetic Calorimeter. Although the latter gives the absolute value of the luminosity, it has a long response time (updated every 30s). By contrast, the SSAL has a much shorter response time (updated every 1s). Moreover, it takes a normalization of the luminosity with respect to the beam current,

$$L_{SpeLumP} = \frac{L_{ZeroLumP} / N_{bCollide}}{(I_{BER} / N_{bBER}) * (I_{BPR} / N_{bBPR})} \quad , \tag{4}$$

where $L_{ZeroLumP}$ is the small-angle luminosity given by a zero-degree detector [15], $N_{bCollide}$ is the number of collision bunches, $I_{BER}$ and $I_{BPR}$ are the electron and positron beam current respectively, and $N_{bBER}$ and $N_{bBPR}$ are the number of electron bunches and positron bunches respectively. As a result, the decay of the beam current during the RCDS scan has a very weak effect on the SSAL.

In addition, the time $t_v$ needs to be known, a part of which is taken for the code to calculate the parameter values of the relevant components and to set the parameters to satisfy the variable variations. For the parameter of a specific component, its value is changed step by step with a limit step size, so the time is related to the ranges of the variables. The major part of $t_v$ is taken for the code to wait to obtain a reliable value of the objective function after sending out the instructions of the variable variations. To this end, a series of measurements have been made. The results show that the state of the machine becomes stable again 7 seconds after a change of variables. Thus, to obtain a trust worthy response for the luminosity for a changed variable, it is best to take more than 7 s to complete an evaluation.

In order to ensure the safety and reliability of the online running, some codes are provided to implement the following functionality. Before starting the machine experiment, the program calculates the range of the variables to be tuned according to the real-time running state, and gives warnings for unreasonable sets. Moreover, the program can stop execution if there is a sudden drop in the beam lifetime or beam current during running.

### 3.2 Experimental results

With the preparations mentioned above, experimental studies of the RCDS scan were performed in BEPCII. As a preliminary test of the effectiveness of this method in a collider, only the 4 offset variables were considered, i.e., ($x$, $x'$, $y$, $y'$) at the IP. We set an optical model of BEPCII, which was already found by manual tuning, as the reference state, with SSAL of about 51 mA$^{-2}$. The variables were deliberately set to deviate from the reference and then the RCDS scan was performed to observe the increase in luminosity.

First, right from the reference, we set the offset with a deviation of Ω (as shown in Table 2). The SSAL decreased to be about 17 mA$^{-2}$. The corresponding state of the machine was used as the initial state to run the RCDS algorithm. During the run, getting an evaluation took about 12 seconds, of which about 2 seconds were taken to execute the instructions to change the currents of relevant correctors, 8 seconds were waiting to get a stable luminosity value, and 2 seconds were taken to obtain the objective function value as an average of three readings of the luminosity (updated every 1s). In total, it took about 34 minutes to get 137 evaluations for the whole run. Fig. 4 shows the evolution of the objective function over the duration of the experiment. Finally, the objective was tuned to about −52 mA$^{-2}$ after 2 iterations. The offset turned out to be $\Omega_{opti}$ (as shown in Table 2) relative to the initial state, which is close to −Ω except for the horizontal angular deviation. This is because the luminosity is not so sensitive to the horizontal angular deviation within the present scanning range. During the running, the objective tended to decrease, which indicates the recovery of the luminosity.

Table 2. Comparison of the relative offset and the objective values between the reference setting and the experimental result

| relative offset | $x$ (mm) | $x'$ (mrad) | $y$ (mm) | $y'$ (mrad) | SSAL (mA$^{-2}$) |
|---|---|---|---|---|---|
| reference | 0 | 0 | 0 | 0 | 51 |
| Ω / initial state | −0.6 | 0.15 | 0.002 | 0.2 | 17 |
| $\Omega_{opti}$ | 0.5898 | 0.1911 | −0.0022 | −0.1354 | 52 |

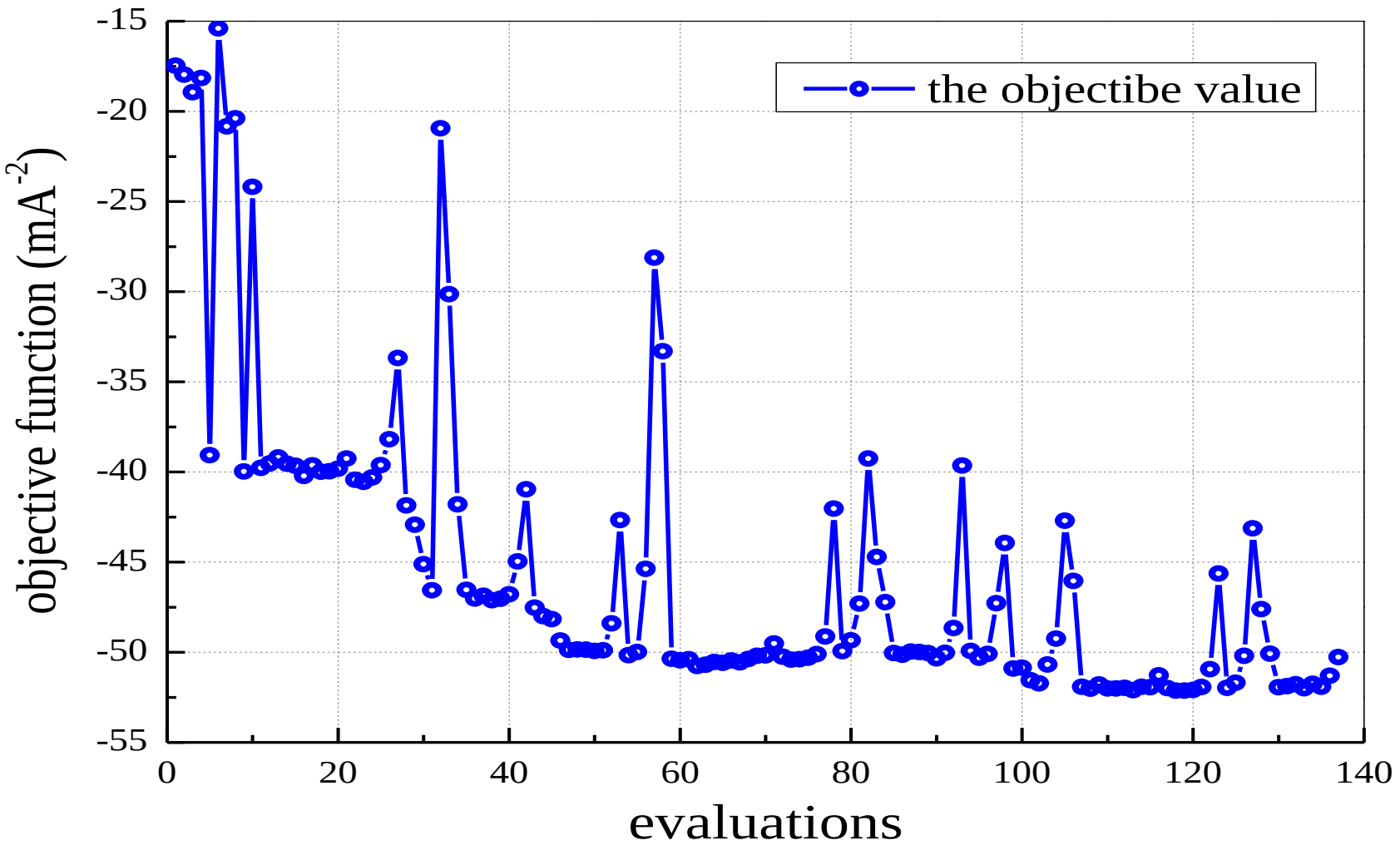

Fig. 4. Evolution of objective function during luminosity tuning online with the RCDS algorithm. The application tuned the luminosity by scanning the variables (*x*, *x'*, *y*, *y'*) sequentially within the first few iterations. About 50 evaluations were taken in every iteration.

Another experiment was then conducted following a similar pattern to that above, except that the variables were scanned with wider ranges. The goal here was to test the experimental effect on dealing with the practical situation in operation, where operators usually have to tune variables within a large range. From experience, the luminosity of BEPCII is sensitive to vertical displacement. So, we deliberately increased the vertical displacement to $\Omega_y$ = 0.025 mm to reduce the reference luminosity. At this point, the SSAL was only 4.3 mA$^{-2}$. We widened the range of vertical displacement from the preceding $4\sigma_y$ to $20\sigma_y$ ($\sigma_y$ = 5 $\mu$m) and ran the RCDS algorithm to tune the luminosity online. In this run, the objective was calculated from an individual reading of the luminosity. About 10 seconds were taken to complete an evaluation, with a total of 138 evaluations taking about 24 minutes. Fig. 5 shows the evolution of the objective function during the luminosity tuning online with the RCDS algorithm. The expected optimal value for the variable of vertical displacement is not found in the first iteration. In the second iteration, when the vertical displacement is found to be close to the negative deviation, $-\Omega_y$, one can see a sudden drop in the objective. This is actually a sudden rise in the SSAL. Finally, the objective is tuned to about −50 mA$^{-2}$ with vertical displacement −0.0273 mm. The evolution of the luminosity recorded by BESIII is shown in Fig. 6, and further verifies the experimental effectiveness.

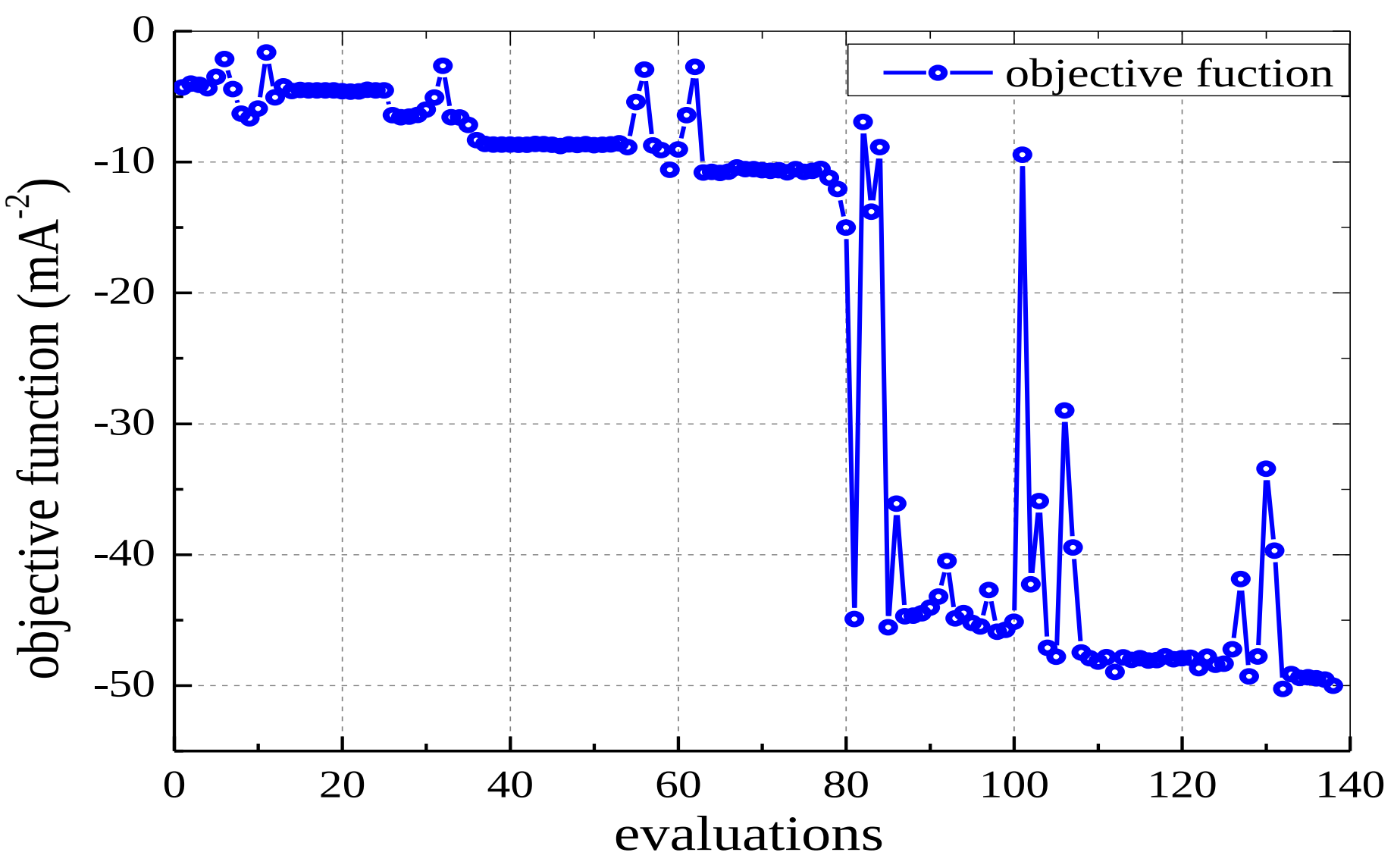

Fig. 5. Evolution of objective function during luminosity tuning online with the RCDS algorithm. The application tuned the luminosity by scanning the variables (*x*, *x'*, *y*, *y'*) sequentially within the first few iterations. About 50 evaluations were taken in every iteration.

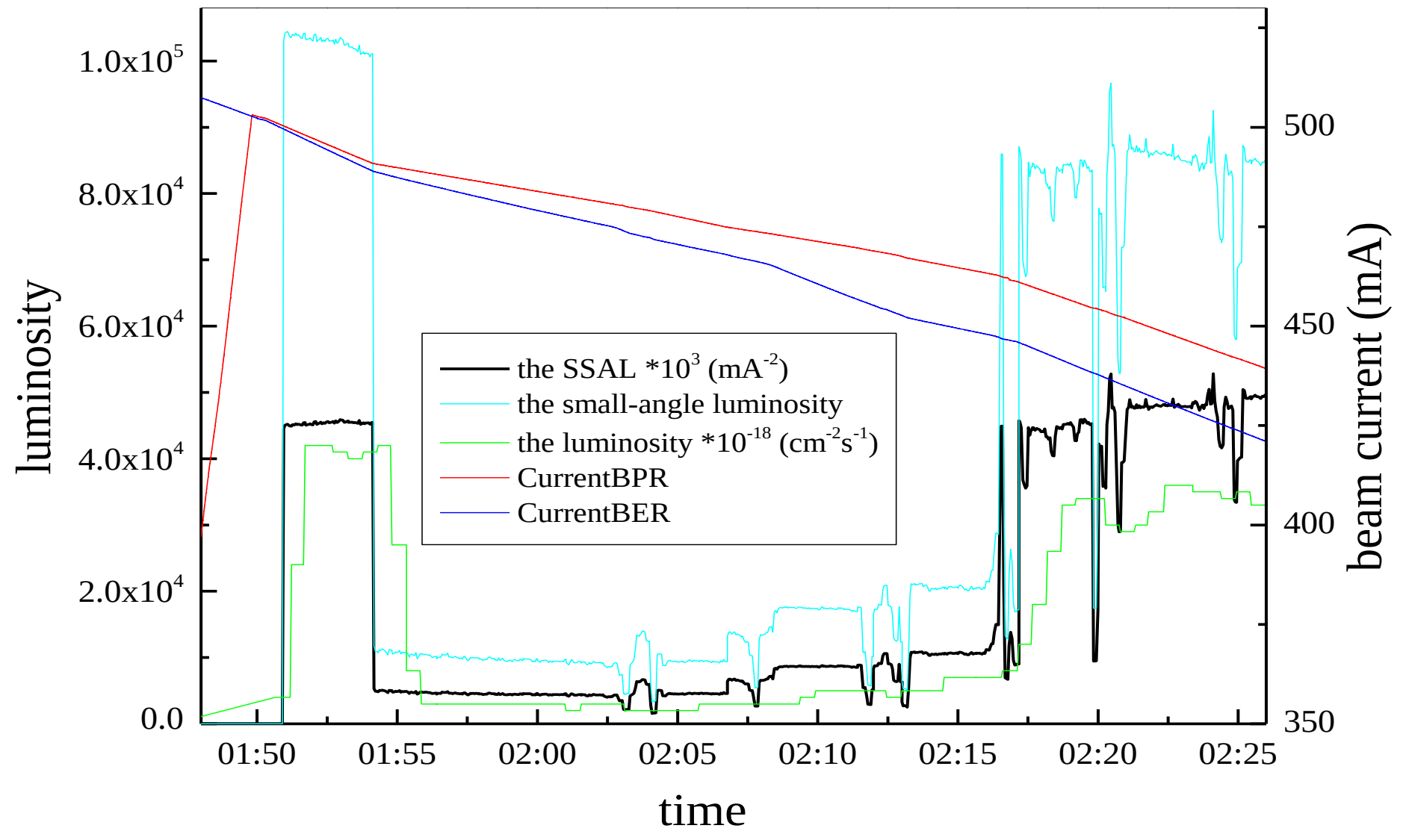

Fig. 6. (color online) Evolution of the luminosity recorded by BESIII. Unlike the recovery of the SSAL, both the small-angle luminosity and the luminosity recorded by BESIII achieve a partial recovery, which results from the effect of the decay of the beam current.

## 4 Discussion

In this paper, we have explored the feasibility of online luminosity tuning using the RCDS method, in numerical simulation, as well as in machine experiments with the BEPCII collider. In the experiments, four variables of orbital offset were used as knobs in the experiments. They were first deviated from the reference standard, which led to a reduced luminosity. An automatic scan of these knobs was then launched with the RCDS method. Convergence was reached within a few iterations, and a recovery of luminosity was also observed.

To implement online optimization of luminosity with RCDS, however, one needs to face the difficulty related to the experiment time. As we know, an actual luminosity optimization (instead of the luminosity recovery experiments shown in this paper) requires delicate tuning of all the related variables. It appears, however, that the minimum time taken for each evaluation is almost fixed, and the total experiment time grows with increasing number of variables. In the presented experiments, only 4 variables were considered and it took about 30 minutes to achieve a high enough luminosity. It can be expected that, if more variables (e.g., 20 variables) are scanned, the time will be too long to achieve a promising result during a collision period (typically ~1h). Thus, it is really hard to finish a scan of all the variables at one time while controlling time to a reasonable level.

As a compromise, another approach can be considered, where one divides the variables into several groups (orbital tuning, working point, coupling factors), and does an RCDS scan in turn for each group, or even in different collision periods, until all the optimal values of the variables are found. This is because different groups of variables are controlled by parameters of different components (for instant, correctors for orbital offset, and skew quadrupoles for coupling factors) and their contributions to luminosity are largely independent. Of course, the equivalence between these two approaches needs to be verified, and the contributions from different groups of variables to luminosity needs to be analyzed. Detailed studies are under way and the results will be presented in a forthcoming paper.